# Structural transitions in superconducting NbTiN thin films

Siddhesh Sanjay Yeram[1#], Sonam Bhakat[1#], Subhashree S. Dash[1], Avradeep Pal[1,2*]

[1]*Department of Metallurgical Engineering and Materials Science, Indian Institute of Technology, Bombay, Mumbai, Maharashtra 400076, India*

[2]*Centre of Excellence in Quantum Information, Computation, Science and Technology, Indian Institute of Technology, Bombay, Mumbai, Maharashtra 400076, India*

**Abstract:** Superconducting NbTiN thin films have garnered extensive interest due to their use in Superconducting Nanowire Single-Photon Detectors (SNSPDs) and other low-temperature applications for potential use in quantum computing and nanoelectronics. This study examines structural phase transitions observed in NbTiN thin films by analyzing the grazing angle x-ray diffraction patterns of a set of reactive magnetron sputter deposited NbTiN thin films with varying nitrogen partial pressures in the reactive gas mixture. The superconducting transition temperature ($T_C$) of the NbTiN thin films showed a correlation with the crystal structure, with the highest $T_C$ of 14.26 K obtained for the highly crystalline FCC phase.

NbTiN, or Niobium Titanium Nitride, a transition metal nitride known for its superconducting properties, is a crucial component in various quantum technologies and photon detection systems. It has been widely used for making superconducting microwave resonators[1], superconducting accelerator cavities[2], SIS mixers[3], superconducting nanowire single-photon detectors (SNSPDs)[4–8], LiDAR[9,10], and cryogenic resistors[11]. NbTiN and NbN have similar electronic, superconducting, and structural properties. However, NbTiN has some benefits over NbN. NbTiN has shown lower resistivity[4,12,13], lower kinetic inductance[4,12,13], and a slightly higher superconducting transition temperature ($T_C$) than NbN[11]. SNSPDs fabricated with NbTiN have shown very high detection efficiency (nearly 99%)[14–16], very low dark counts (< 10 Hz)[17–19], low timing jitter (few picoseconds)[10,14,18,19], and low reset times (few nanoseconds)[13,14,16–18]. These photon detectors have played an important role in quantum communication and quantum computing[19–21].

 NbTiN thin films have been deposited using various deposition methods, including reactive magnetron sputtering[7,18], high-temperature chemical vapor deposition (HTCVD)[22], and atomic layer deposition (ALD)[23]. Of these, reactive DC magnetron sputtering is the principal method of deposition for the NbTiN thin films due to the controlled and high deposition rates, high purity, and uniform deposition[24,25]. The major control parameters for reactive magnetron sputtering are the gas flow rate, reactive gas composition, sputtering power, target-substrate distance, substrate temperature, pressure in the deposition chamber, and deposition time. In this study, we aim to investigate the effect of varying nitrogen percentages in the reactive gas mixture on the crystal structure and superconducting transition temperature of the NbTiN thin films, keeping all other parameters constant.[*]

Transition metal nitrides are known to have complex crystal structures, and understanding them plays a pivotal role in enhancing their electronic, structural, magnetic, and superconducting properties for usage in various applications. Various studies on effect of different substrates[7,26], Nb and Ti percentage[8,22,23,27], varying sputtering power, and pressure[3,8,25,28], thickness[3,28], and nitrogen percentage[28,29] in the gas mixture on the $T_C$ of NbTiN films have been conducted, but these do not take into account the crystal structure of the thin films deposited and how that affects the $T_C$. In this paper, we take a deeper dive into the changes in the crystal of the NbTiN thin films with varying nitrogen content, particularly at very low and very high nitrogen percentages, different phases of NbTiN observed, how it affects the $T_C$ of the thin films, and the amorphization of the films to obtain the required properties for use in various superconducting devices including SNSPDs.

---

[*] *avradeep@iitb.ac.in*
#*Both authors contributed equally to this work*

**Experimental Methods:**

| Sample name | N$_2$% | Power (Watts) | Voltage (Volts) | Current (mA) |
|---|---|---|---|---|
| 94E | 10 | 75 | 549 | 136 |
| 94F | 30 | 65 | 536 | 121 |
| 102A | 10 | 65 | 602 | 107 |
| 102B | 20 | 65 | 585 | 111 |
| 102C | 30 | 65 | 636 | 101 |
| 102D | 40 | 65 | 633 | 103 |
| 126A | 5 | 65 | 478 | 135 |
| 126B | 15 | 65 | 497 | 130 |
| 126C | 20 | 65 | 500 | 129 |
| 126D | 25 | 65 | 501 | 128 |
| 141A | 2.5 | 65 | 472 | 137 |
| 141B | 12.5 | 65 | 481 | 134 |
| 141C | 17.5 | 65 | 485 | 132 |
| 141D | 50 | 65 | 541 | 119 |
| 160B | 1.5 | 65 | 290 | 224 |
| 160C | 3.8 | 65 | 286 | 225 |

**Table 1.** Conditions used for growth of NbTiN thin films deposition

NbTiN thin films were deposited with DC magnetron sputtering using a custom-design Ultra-High Vacuum Multilayer Magnetron Sputtering System. The target used was a Niobium-Titanium (Nb-Ti 69-31 atomic percent) rectangular target with 99.9% purity. The distance between the target and substrate was 3.5 cm. Substrates used were [100] oriented SiO$_2$ on n-doped Si. The pressure in the deposition chamber was maintained at 1.5 ± 0.01 Pa, and the base pressure before starting the deposition was in the order of 10$^{-7}$ Pa. The nitrogen percentage in the gas mixture is given by (N$_2$ pressure/ Total pressure) in the gas mixing unit. NbTiN thin films with four different nitrogen percentages were deposited in the same run by placing samples on a rotating disk in the flange, using a masking plate under the NbTi target, and the rotating motion was carried out using a computer-controlled stepper motor. Thin films were deposited with 13 different nitrogen percentages (1.5, 2.5, 3.8, 5, 10, 12.5, 15, 17.5, 20, 25, 30, 40, and 50) across four different deposition runs, and the details of samples in each run are listed in Table 1. The superconducting transition temperature ($T_C$) and in-plane critical field of the thin films were measured using a liquid cryogen-free closed-cycle cryostat. The $T_C$ is considered as the temperature corresponding to 50 percent of the maximum resistance on the resistance vs temperature graph. The transition width is considered as the temperature difference between temperatures corresponding to 10 percent and 90 percent of the maximum resistance. Grazing Incidence X-Ray Diffraction (GIXRD) was used to get XRD plots for the NbTiN thin films.

**Results and Discussion:**

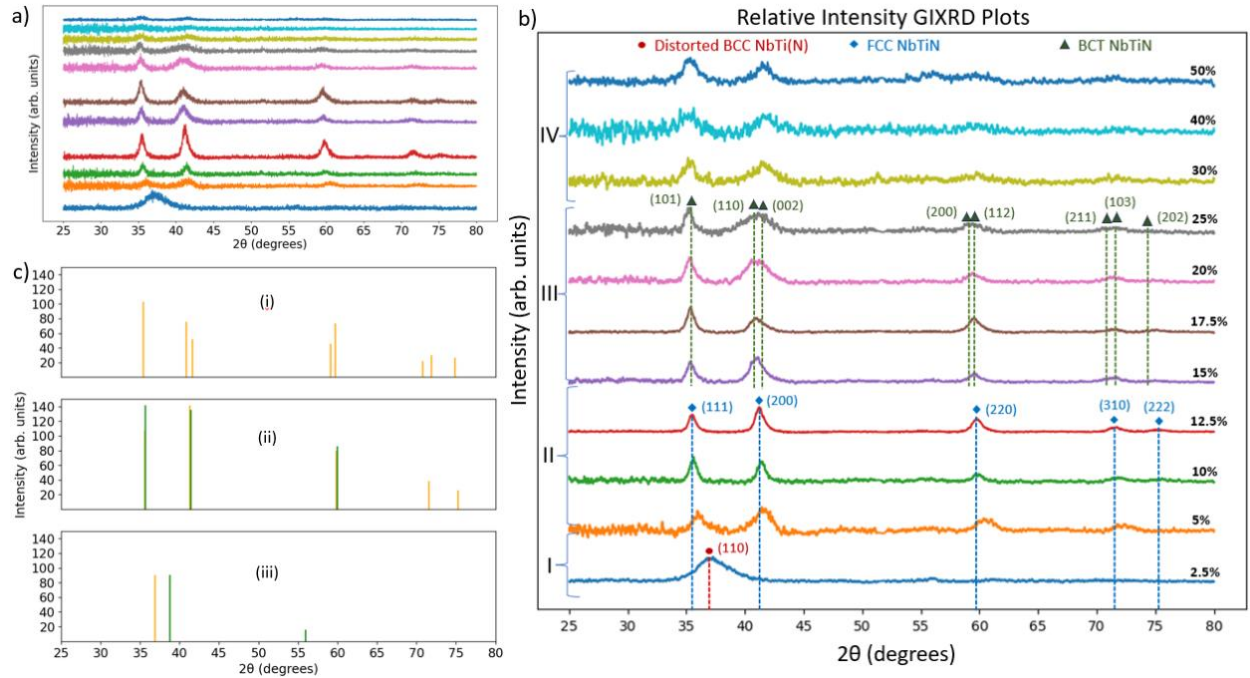

**Figure 1**. a) Grazing incidence x-ray diffraction (GIXRD) pattern of NbTiN thin films grown with different nitrogen pressures (without the substrate peaks), b) relative intensity GIXRD plots of the NbTiN films showing peaks corresponding to the different phases, c) peak positions for the (i) BCT, (ii) FCC, and (iii) BCC phase, yellow lines show the peaks obtained from XRD while green lines show peaks obtained from crystallography open database[27,30].

On the introduction of nitrogen gas at low pressures in transition metals, it has been shown that nitrogen atoms prefer to occupy the interstitial sites, leading to expansion and distortion of the unit cell. At moderate nitrogen pressure, it favors the formation of metal nitride compounds, and at higher nitrogen pressure, the structure gets distorted, leading to the formation of an amorphous phase[31–33]. We observed a similar trend in the NbTiN thin films. NbTi has a body-centered cubic (BCC) structure ($Im\bar{3}m$ space group) with lattice parameter $a_{NbTi}$ = 3.286 Å[30]. When nitrogen is introduced in this structure at a lower concentration (Fig. 1.b-region I), it prefers to occupy the interstitial sites. It has been shown that transition metals and alloys having BCC structure have non-regular tetrahedral voids (Wyckoff 8c sites) with $r_{int}$ = 0.291*$r_{BCC}$ (where $r_{int}$ denotes the maximum radius atom or ion that can be accommodated in the respective void sites) and non-regular octahedral voids (6b sites) with $r_{int}$ = 0.155*$r_{BCC}$ (two atoms are closer than other neighboring atoms)[34]. Calculating $r_{BCC}$ for NbTi unit cell using the well-known expression for BCC unit cell $\sqrt{3}$* $a_{NbTi}$ = 4* $r_{BCC}$, we get $r_{BCC}$ = 1.423 Å. Using this information, we get the radius of interstitial voids as 0.414 Å and 0.221 Å for non-regular tetrahedral voids (8c sites) and non-regular octahedral voids (6b sites), respectively. These are significantly less than the radius of the nitrogen atom 0.65 Å[35]. Thus, for accommodating the nitrogen atoms, the structure has to distort significantly. This can be seen from the broadening and shifting of the (110) peak of the NbTi BCC structure (Fig. 1.a and

1.c). Although the non-regular tetrahedral voids (8c sites) in a BCC lattice exhibit a larger interstitial volume, empirical evidence indicates a preference for nitrogen atoms to occupy octahedral interstices (6b sites) as shown in Figure 2.b.I. This preference is attributed to the octahedral sites being more conducive to strain alleviation, facilitated through the displacement of two adjacent atoms. In contrast, the accommodation of nitrogen atoms in non-regular tetrahedral voids (8c sites) would necessitate a higher strain energy, making it a less favorable site[36]. As the nitrogen percentage is increased, we can see an increase in the lattice parameter of the BCC unit cell from 3.286 Å to 3.693 Å (Fig. 2.a.i).

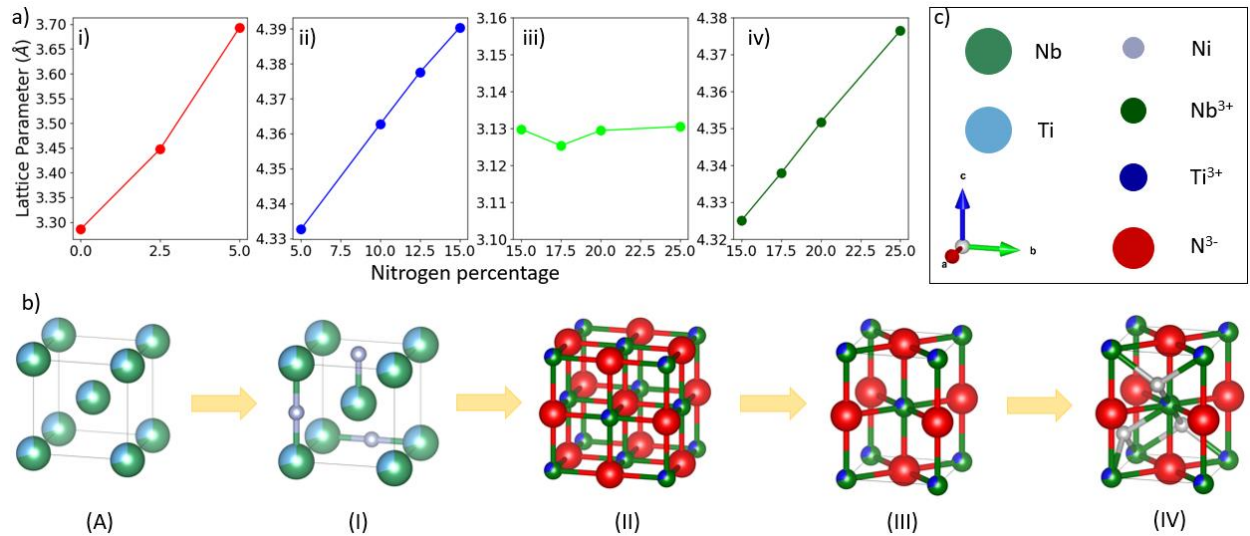

**Figure 2**. a) Variation of lattice parameters for different nitrogen percentages, where i) is for the BCC phase, ii) is for the FCC phase, and (iii), (iv) correspond to lattice parameters a and c, respectively, for the BCT phase, b) A schematic representation of the change in crystal structure of NbTiN with change in nitrogen &, (A) corresponds to pure NbTi, (I) corresponds to low nitrogen percentage, (II) is FCC NbTiN, (III) is the BCT phase, and (IV) corresponds to high nitrogen percentage (visualized using VESTA software), c) The viewing direction for the unit cells and the color coding used for representing various atoms and ions in the crystal structures.

In the moderate nitrogen percentage range (Fig. 1.b-region II), we can see the formation of the face-centered cubic (FCC) NbTiN phase, which can be seen by the formation of sharp peaks matching with the ones obtained from the literature (Fig. 1.c.ii)[27]. At 5% nitrogen, we get peaks corresponding to both the BCC and the FCC phases. As the nitrogen percentage in the reactive gas increases from 5% to 12.5%, we can see the formation of the FCC phase with the peaks becoming sharper, achieving the highest sharpness and, hence, the highest degree of crystallinity at 12.5%, suggesting that stochiometric NbTiN FCC phase is formed around 12.5%. The NbTiN FCC phase shows a NaCl-like crystal structure ($Fm\bar{3}m$ space group)[27,37,38] with nitrogen ions occupying the octahedral voids (4b) (Fig. 2.b.II), and the lattice parameter for FCC phase ($a_{NbTiN}^{FCC}$) obtained from analyzing the XRD peaks for 12.5% was 4.378 Å. The edge length (or the lattice parameter) of the FCC unit cell can also be calculated by the expression a = 2*($r_{anion}$+$r_{cation}$). Using the data for ionic radii of $N^{3-}$ ($r_N$)[39], $Nb^{3+}$ ($r_{Nb}$)[39], and $Ti^{3+}$ ($r_{Ti}$)[39] and taking $r_{cation}$ = 0.7* $r_{Nb}$ + 0.3* $r_{Ti}$, we get $a_{calculated}$ = 4.33 Å, which is very close to the lattice parameter obtained experimentally, further confirming the formation of an ionic compound.

On further increase in the nitrogen percentage from 15% to 25% (Fig. 1.b-region III), multiple peaks were observed near 41 and 59 degrees in the XRD plots. Furthermore, the lattice parameters, determined using the expression $a = \frac{\lambda * \sqrt{h^2 + k^2 + l^2}}{2 * \sin\theta}$ (considering a cubic unit cell) for different 2θ values in the same XRD plot, showed significant discrepancies, indicating that the unit cell might not conform to the FCC structure at these nitrogen percentages. This discrepancy is demonstrated in Table 2.

| %N$_2$ | Peak 1 | Peak 2 | Peak 3 | Peak 4 | a1 | a2 | a3 | a4 | σ |
|---|---|---|---|---|---|---|---|---|---|
| 10 | 35.59(9) | 41.41(5) | 59.86(0) | 71.74(9) | 4.365 | 4.357 | 4.367 | 4.360 | 0.0045 |
| 12.5 | 35.50(8) | 41.14(5) | 59.76(4) | 71.57(6) | 4.375 | 4.384 | 4.373 | 4.369 | 0.0065 |
| 15 | 35.37(2) | 41.06(8) | 59.55(7) | 71.41(8) | 4.392 | 4.392 | 4.387 | 4.377 | 0.0070 |
| 17.5 | 35.36(9) | 40.89(2) | 59.54(4) | 71.51(7) | 4.392 | 4.410 | 4.388 | 4.371 | 0.0158 |
| 20 | 35.29(8) | 41.46(8) | 59.43(5) | 71.33(7) | 4.401 | 4.352 | 4.395 | 4.382 | 0.0219 |
| 25 | 35.21(0) | 41.22(9) | 59.33(2) | 71.64(7) | 4.411 | 4.376 | 4.402 | 4.365 | 0.0217 |

**Table 2.** 2θ values (in degrees), corresponding calculated lattice parameters a1, a2, a3, and a4 (in angstrom), and standard deviation of the lattice parameter values considering an FCC phase for different nitrogen percentages.

| hkl | 15% N$_2$ Obtained peaks | 15% N$_2$ calculated peaks | 17.5% N$_2$ Obtained peaks | 17.5% N$_2$ calculated peaks | 20% N$_2$ Obtained peaks | 20% N$_2$ calculated peaks | 25% N$_2$ Obtained peaks | 25% N$_2$ calculated peaks |
|---|---|---|---|---|---|---|---|---|
| 101 | 35.37(2) | 35.37(2) | 35.36(9) | 35.36(9) | 35.29(8) | 35.29(8) | 35.22(0) | 35.22(0) |
| 110 | 40.56(4) | 40.73(8) | 40.41(8) | 40.79(8) | 40.50(4) | 40.74(3) | | 40.72(8) |
| 002 | 41.73(4) | 41.73(4) | 41.60(5) | 41.60(5) | 41.46(8) | 41.46(8) | | 41.22(2) |
| 200 | | 58.97(6) | | 59.06(7) | | 58.98(3) | | 58.96(0) |
| 112 | 59.55(3) | 59.73(9) | 59.54(4) | 59.68(5) | 59.43(5) | 59.53(7) | 59.33(8) | 59.33(7) |
| 211 | | 70.68(2) | | 70.76(3) | | 70.64(3) | | 70.57(5) |
| 103 | 71.41(8) | 72.06(8) | 71.51(7) | 71.88(5) | 71.33(7) | 71.65(0) | 71.65(1) | 71.26(0) |
| 202 | 74.93(9) | 74.83(2) | 74.79(6) | 74.82(5) | | 74.65(5) | | 74.46(7) |

**Table 3.** 2θ values (in degrees) for different hkl values obtained from XRD plots and from calculations considering a BCT phase for different nitrogen percentages

We observe that with the increase in nitrogen percentage, we get a significant increase in the standard deviation (σ) of the lattice parameter values obtained from the different peaks considering an FCC phase. The above discrepancy is suggestive of a structural transition in NbTiN with the change in $N_2$ partial pressure of reactive gas. To account for this, we take note of the well-known FCC to body-centered tetragonal (BCT) transition[40,41]. Considering this phase transition, we indexed the first three peaks in the XRD plot (corresponding to 2θ values near 35, 40, and 41 degrees) to the first three peaks for a BCT structure with hkl values (101), (110), and (200). Using the two highest intensity peaks (which corresponded to the hkl values (101) and (200)), we calculated the lattice parameters $a^{BCT}_{NbTiN}$ and $c^{BCT}_{NbTiN}$ for the BCT unit cell using the expressions $n\lambda = 2d\sin\theta$ and $\frac{1}{d^2} = \frac{h^2 + k^2}{a^2} + \frac{l^2}{c^2}$ by substituting the appropriate θ, h, k, and l values and solving the simultaneous equations in two variables a and c. Using the obtained $a^{BCT}_{NbTiN}$ and $c^{FCC}_{NbTiN}$ values, we calculated the 2θ values corresponding to the other BCT structure peaks using the above-mentioned expressions and the reflection conditions for a BCT unit cell. The 2θ values obtained from the above calculations closely matched the 2θ values obtained from the XRD plot (as shown in Table 3), thus confirming a phase transition from the FCC phase to a BCT phase. Details pertaining to these calculations for different nitrogen percentages are shown in Table 3.

It is found that $a^{BCT}_{NbTiN}$ remains constant (around 3.129 Å, Fig. 2.a.iii) while $c^{FCC}_{NbTiN}$ increases from 4.325 Å to 4.376 Å (Fig. 2.a.iv) with an increase in nitrogen percentage from 15 to 25. On increasing the nitrogen content above 25%, we can see a sharp decrease in the intensity of the peaks obtained (Fig. 1.a), which suggests that the thin films have transitioned into an amorphous phase. This amorphization is desirable for the fabrication of SNSPDs[5] as it leads to the formation of thin films with greater surface smoothness and isotropy[31,42] due to the lack of grains and grain boundaries.

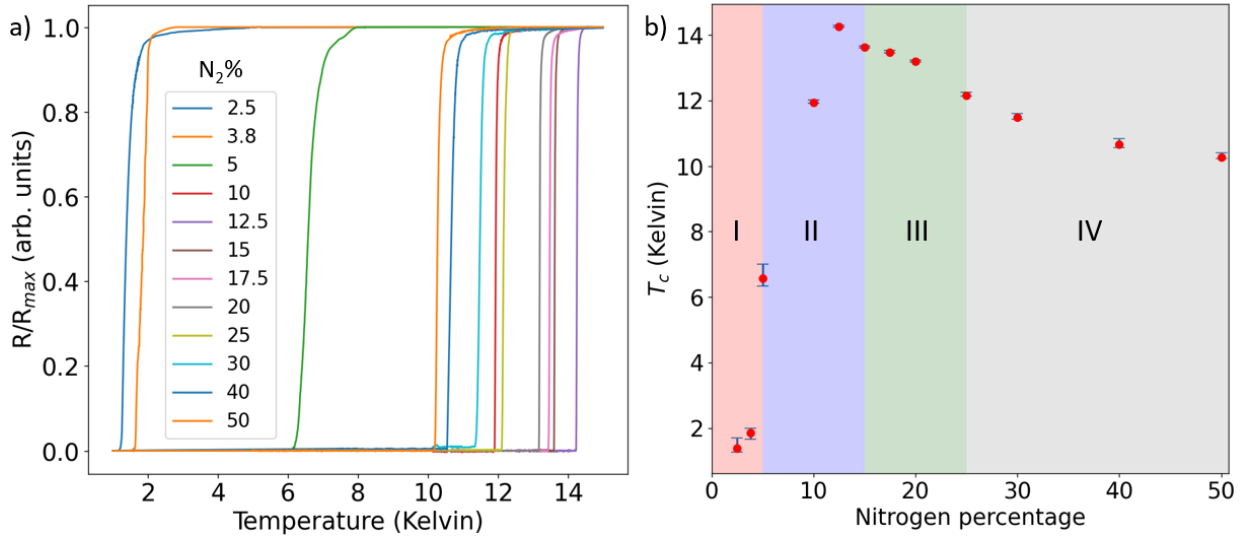

**Figure 3**. a) Superconducting transition temperatures ($T_C$) of the NbTiN thin films deposited at different nitrogen percentages, b) $T_c$ (red dot) with transition width (blue lines) as a function of nitrogen percentage. I, II, III, and IV represent BCC, FCC, BCT, and amorphous regions, respectively.

The trend of superconducting transition temperature ($T_C$) with nitrogen percentage can be seen in Figure 3.b. NbTi has a $T_C$ of 10 K, and introducing even a small amount of nitrogen into the structure significantly reduces the $T_C$ to 1.38 K (for 2.5% nitrogen), suggesting that the distortion of the structure has a very significant effect on the $T_C$ (region I in Figure 3b). As the FCC phase starts forming, the $T_C$ of the thin films starts increasing sharply, increasing from 6.57 K to 14.26 K as the nitrogen percentage is increased from 5% to 12.5% (region II in Figure 3b). On further increase in nitrogen percentage, in the BCT range, the $T_C$ reduces in comparison to the FCC phase but remains nearly constant (region III in Figure 3b). Observations indicate a progressive decline in the critical temperature ($T_C$) beyond a nitrogen concentration of 20%, stabilizing in the range of 40% to 50%. Notably, NbTiN thin films with pronounced and intense peaks, particularly in the 10% to 25% nitrogen concentration range, exhibit the highest $T_C$ width narrow transition widths (<0.1 K). This pattern suggests a significant correlation between $T_C$ and the crystallinity and crystal structure of the films. More studies need to be done to get a deeper understanding of these correlations.

In conclusion, this study shows that the percentage of nitrogen in the reactive sputtering gas mixture has a significant impact on the crystal structure and $T_C$ of the NbTiN thin films. On increasing the nitrogen percentage in the gas mixture, we first form a distorted BCC phase, followed by the formation of an FCC phase, which then transforms into a BCT phase upon further increasing the nitrogen percentage. Finally, we get amorphous NbTiN thin films at a very high nitrogen percentage. The $T_C$ of the thin films also shows a correlation with the crystal structure of the thin films, showing high values corresponding to the highly crystalline FCC phase and very low values for the distorted BCC phase. The comparatively high $T_C$ and amorphous nature obtained at high nitrogen percentage thin films makes them ideal for use in SNSPDs[5], which require high film uniformity, and the thin films fabricated with very low percentages of nitrogen are great potential candidates for the fabrication of cryogenic resistors[11]. This work paves the way for further studies to better understand and utilize NbTiN in various superconducting and low-temperature applications.